\documentclass{SciPost}

\hypersetup{
    colorlinks,
    linkcolor={red!50!black},
    citecolor={blue!50!black},
    urlcolor={blue!80!black}
}

\usepackage[bitstream-charter]{mathdesign}
\DeclareSymbolFont{usualmathcal}{OMS}{cmsy}{m}{n}
\DeclareSymbolFontAlphabet{\mathcal}{usualmathcal}

\fancypagestyle{SPstyle}{
\fancyhf{}
\lhead{\colorbox{scipostblue}{\bf \color{white} ~SciPost Physics Proceedings }}
\rhead{{\bf \color{scipostdeepblue} ~Submission }}

\fancyfoot[C]{\textbf{\thepage}}
}

\usepackage{graphicx}        
\usepackage{rotating}
\usepackage{geometry} 
\usepackage{epsfig}
\DeclareGraphicsExtensions{.pdf,.png,.jpg}

\begin{document}

\pagestyle{SPstyle}

\begin{center}{\Large \textbf{\color{scipostdeepblue}{
Investigating the hadron nature of high-energy photons with PeVatrons\\
}}}\end{center}

\begin{center}\textbf{
Giuseppe Di Sciascio
}\end{center}

\begin{center}
INFN - Roma Tor Vergata, Roma, Italy 
\\
 \href{mailto:giuseppe.disciascio@roma2.infn.it}{\small giuseppe.disciascio@roma2.infn.it}
\end{center}

\definecolor{palegray}{gray}{0.95}
\begin{center}
\colorbox{palegray}{
  \begin{tabular}{rr}
  \begin{minipage}{0.36\textwidth}
    \includegraphics[width=55mm]{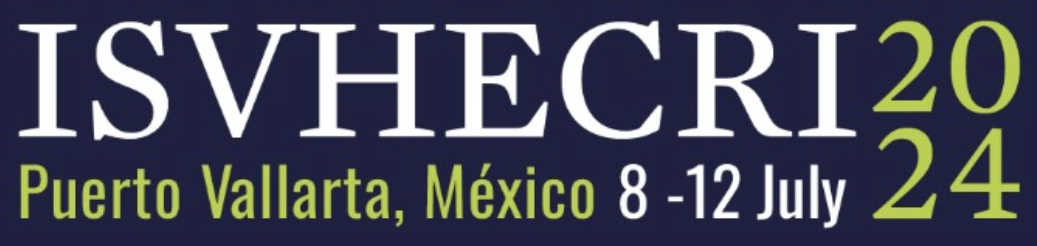}
  \end{minipage}
  &
  \begin{minipage}{0.55\textwidth}
    \begin{center} \hspace{5pt}
    {\it 22nd International Symposium on Very High \\Energy Cosmic Ray Interactions (ISVHECRI 2024)} \\
    {\it Puerto Vallarta, Mexico, 8-12 July 2024} \\
    \doi{10.21468/SciPostPhysProc.?}\\
    \end{center}
  \end{minipage}
\end{tabular}
}
\end{center}

\section*{Abstract}
{\bf

In high energy Gamma-Ray Astronomy with shower arrays the most discriminating signature of the photon-induced showers against the background of hadron-induced cosmic-ray is the content of muons in the observed events. 
In the electromagnetic $\gamma$-showers the muon production is mainly due to the photo-production of pions followed by the decay $\pi\to\mu\nu$, with a small relative probability of order $\alpha$ ($\simeq$ 1/137).
Therefore, the number of muons is typically a few percent of that in a hadron showers where muons are abundantly generated by charged pions decay.\\
In high energy photo-production process the photon exhibits an internal structure which is very similar to that of hadrons.
Indeed, photon-hadron interactions can be understood if the physical photon is viewed as a superposition of a bare photon and an accompanying small hadronic component which feels conventional hadronic interactions.\\
Information on photo-production $\gamma$p and $\gamma\gamma$ cross-sections are limited to $\sqrt{s}\leq$ 200 GeV from data collected at HERA. Starting from  $E_{lab}\approx$100 TeV the difference between different extrapolations of the cross sections increases to more than 50\% at $E_{lab}\approx$10$^{19}$ eV, with important impact on a number of shower observables and on the selection of the photon-initiated air showers. \\
Recently, the LHAASO experiment opened the PeV-sky to observations detecting 40 PeVatrons in a background-free regime starting from about $E_{lab}\approx$ 100 TeV. This result provides a beam of pure high energy primary photons allowing to measure for the first time the photo-production cross section even at energies not explored yet.
The future air shower array SWGO in the Southern Hemisphere, where the existence of Super-Pevatrons emitting photons well above the PeV is expected, could extend the study of the hadron nature of the photons in the PeV region.\\
In this contribution the opportunity for a measurement of the photo-production cross section with air shower arrays is presented and discussed.
}

\vspace{\baselineskip}

\noindent\textcolor{white!90!black}{%
\fbox{\parbox{0.975\linewidth}{%
\textcolor{white!40!black}{\begin{tabular}{lr}%
  \begin{minipage}{0.6\textwidth}%
    {\small Copyright attribution to authors. \newline
    This work is a submission to SciPost Phys. Proc. \newline
    License information to appear upon publication. \newline
    Publication information to appear upon publication.}
  \end{minipage} & \begin{minipage}{0.4\textwidth}
    {\small Received Date \newline Accepted Date \newline Published Date}%
  \end{minipage}
\end{tabular}}
}}
}




\section{Introduction}
\label{sec:intro}

The measurement of the muon content in Extensive Air Showers (EAS) is the most powerful method to select $\gamma$-induced events and to reject the background of charged cosmic rays (CR) in ground-based gamma-ray astronomy with air shower arrays.

In the electromagnetic $\gamma$-showers the muon production is due to the dominant channels: photo-production of pions followed by the decay $\pi\to\mu\nu$, prompt leptonic decay of charmed particles in the shower, and electromagnetic pair production $\gamma\to\mu^+\mu^-$.
The main process is the photo-production (PhP) of pions with a small relative probability of order $\alpha$ ($\simeq$ 1/137).
Therefore, the number of muons is typically a few percent of that in a hadron showers where muons are abundantly generated by charged pions decay.

The knowledge of the PhP cross section is therefore crucial to evaluate the expected number of muons in gamma showers and to set a correct threshold to discriminate, in high energy gamma-ray astronomy, the showers induced by the photons from the background due to charged cosmic rays.

As will be discussed in the following, information on PhP $\gamma$p and $\gamma\gamma$ cross-sections are limited to $\sqrt{s}\leq$ 200 GeV from data collected at HERA \cite{hera}.
Starting from  $E_{lab}\approx$ 100 TeV the difference between different extrapolations of the cross sections increases to more than 50\% at $E_{lab}\approx$ 10$^{19}$ eV, with important impact in the value of different observables in EAS \cite{cornet2015}.

The recent observation of about 40 gamma sources above 100 TeV, in a nearly background-free regime, offers, for the first time, the possibility to study the PhP cross section even at energies not explored yet.

\section{The hadronic cross sections}

The photon is the gauge boson of quantum electrodynamics and is regarded as point-like and structureless.
Therefore, its interaction with matter is believed to be entirely electromagnetic, that is, it appears to involve only the charges and magnetic fields of the target particles but not their possible nuclear ("hadronic", "strong") interactions. 

With increasing energy another important manifestation of the photon becomes dominant.
In high energy photo-production, a process where something is produced by the interaction of a high energy photon with hadrons, something like $\gamma + N\to\pi + N$, $\pi\to\mu$, the photon exhibits an internal structure which is very similar to that of hadrons, with a small relative probability of order $\alpha$ ($\simeq$ 1/137).
That is, its interaction cross-sections behave (apart from a normalization factor) very much like hadronic cross-sections, and at the highest energies the photon even appears to 'contain' quarks and gluons, just as a proton. 

The simplest model for describing the hadron nature of the photon is the \emph{Vector Meson Dominance (VMD)} model \cite{vmd}. 
In this model, the photon is assumed to transform, before an interaction, to a neutral vector meson $V$ (such as the $\rho^0$, $\omega$, or $\phi$),  $\gamma\leftrightarrow V$, while the interaction of the bare photon with hadrons becomes negligible at high energies. 
The quark model predicts that the photon should behave as if it were 75\% $\rho$, 8\% $\omega$, and 17\% $\phi$. Thus the $\rho$ is the most important of the vector mesons in mediating photon-hadron interactions.
All vector mesons produced by virtual photons decay immediately and must be observed indirectly through the long-lived particles into which it decays. 

The study of hadronic cross sections and the understanding of their energy dependence is always an important issue in the study of strong interactions (for a comprehensive review see \cite{pancheri}). The hadronic elastic and total cross sections are crucial instruments to probe the so-called soft part of QCD physics, where quarks and gluons are confined. As the energy increases, the total cross section also probes the transition into hard scattering describable with perturbative QCD, the so-called mini-jet region.

Measurements of the hadronic cross sections are made with different techniques due to the different projectiles and targets used. The study of the interactions of primary CR particles with the nuclei of atmosphere allowed to measure the p-air and pp cross sections up to $\sqrt{s}$ = 57 TeV \cite{auger}. Information on photo-production $\gamma$p and $\gamma\gamma$ cross-sections are instead limited to $\sqrt{s}\leq$ 200 GeV from data collected at HERA \cite{hera}.

\begin{figure}[t]
\vfill  \begin{minipage}[t]{.48\linewidth}
  \begin{center}
    \mbox{\epsfig{file=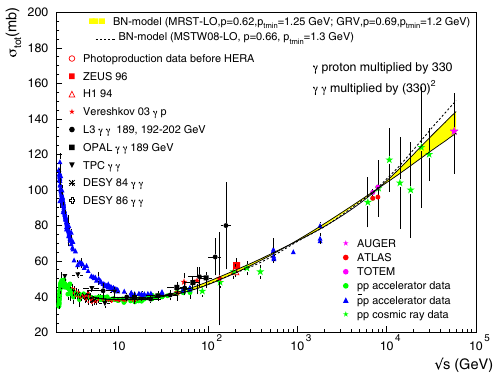,width=6cm}}
  \vspace{-0.5pc}
  \end{center}
\end{minipage}\hfill
\hspace{-0.5cm}
\begin{minipage}[t]{.47\linewidth}
  \begin{center}
    \mbox{\epsfig{file=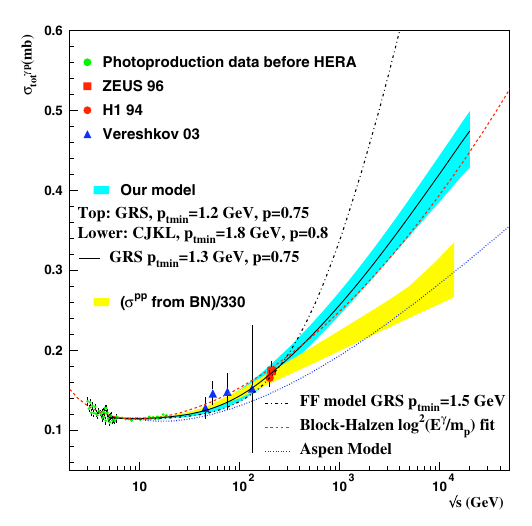,width=6cm}}
  \vspace{-0.5pc}
  \end{center}
\end{minipage}\hfill
\caption{Left panel: Total cross section data for $pp$ and $p\overline{p}$ scattering together with normalized $\gamma$p and $\gamma\gamma$ data. Curves describe predictions from a mini-jet model with soft gluon resummation.\
Right panel: Total $\gamma$p cross sections measured in different experiments compared with expectations from different models (see ref. \cite{pancheri,godbole2009} for references and details).} 
  \label{fig:total-crossect}
\end{figure}

According to experimental results, all total cross sections rise asymptotically with energy, as first observed in analysis of CR data in the early '70. 
In the left panel of Fig. \ref{fig:total-crossect}  a compilation of total pp/p$\overline{\text{p}}$, $\gamma$p and $\gamma\gamma$ cross sections, from accelerators and CR experiments, is shown together to highlight their common features \cite{pancheri}. The dashed and full curves over imposed to the data are obtained from a mini-jet model described in \cite{pancheri,godbole2009}. Since the data span an energy range of four orders of magnitude, with the cross sections in the milli-barn range for proton-proton, micro barn range for PhP and nano-barns for $\gamma-\gamma$, to plot them all on the same scale we must use a normalization factor by multiplying the $\gamma$-p cross section by a factor $\approx$330 and the $\gamma-\gamma$ by (330)$^2$ \cite{pancheri}. 

To describe the $\gamma$-proton interaction, and to calculate the PhP cross-section extrapolated up to the highest CR energies, several models have been developed. They include
\begin{itemize}
\item factorisation models, in which by means of a simple multiplicative factor the photon processes are compared with each other and with the pure proton ones
\item microscopic models, such as Block-Nordsiek models, with quarks and gluons \cite{pancheri,godbole2009}.
\end{itemize}

In the factorization models a brute force factorization is applied to models describing the proton-proton interaction to obtain the PhP cross sections. The photon is fundamentally an electromagnetic object that makes occasional transitions to a hadronic state. 

All factorisation models imply that there is a universal behaviour of the energy dependence, not only at low energy, where one can assume that the hadronic interactions of the photons are those of a vector meson, but also at high energy. In the low energy region PhP has conventionally been calculated \'a la VMD with $\gamma\to\rho\to\pi\pi$. The photon is considered (always) hadron-like and $\sigma_{tot}^{\gamma p}$ = $R_\gamma\cdot \sigma_{tot}^{pp}$, where $R_\gamma$ is the probability that the photon makes occasional transitions to a hadronic state. 

%

The VMD model allows to estimate the factor $R_\gamma$  at $\sqrt{s}$ = 10--20 GeV, before the beginning of the high-energy rise of cross sections. With the $\rho$--meson data $R_\gamma\approx$ 1/360 is obtained, consistent with the normalization factors applied in the left panel of Fig.  \ref{fig:total-crossect}. But the behaviour to be expected at the high end of CR energies cannot be gauged from HERA analyses and \emph{new measurements of the energy dependence of the photon–proton total cross section are needed at higher energies}. 

In the right panel of Fig. \ref{fig:total-crossect} the total $\gamma$-p cross section measured in different experiments is shown compared with expectations from different models (for a detailed discussion see \cite{pancheri,godbole2009}). While at moderate, HERA-like energies, all the models, factorizations and microscopic, give good fits to the data, a remarkable difference between their high energy extrapolations can be appreciated starting from $\sqrt{s}\approx$ 300 GeV. This large difference (more than 50\% at $E_{lab}\approx$ 10$^{19}$ eV) may impact strongly on high-energy CR physics and in particular on the evaluation of the photon content in the primary flux up to the energies investigated by the AUGER experiment. But may also impact on the sensitivity of gamma-ray telescopes, on the determination of the flux and of the cut-off energies in the spectra of gamma sources above 100 TeV, and in the calculation of neutrino flux from astrophysical sources.

The crucial question of factorization to be addressed is following, \emph{is a photon like a proton just multiplied by a constant factor?} We can explore the effects of the hadronic structure of the photon through the analysis of the total cross sections involving photons. The measurement of the $\gamma$p cross section above $\sqrt{s}\approx$ 300 GeV is crucial to disentangle between different models and this goal can be reached only by using CR data.

\section{Pion photo-production in Extensive Air Showers}

Showers induced by photons are primarily electromagnetic cascades powered by pair production and breemstrahlung processes. Occasionally, with a probability of order $\alpha$ (1/137), a photon interact hadronically, producing a sub shower that is essentially hadronic in nature with a normal hadronic muon content. 

The total number of \emph{normal muons} is related to the elemental composition of the CR primary flux and to the characteristics of hadronic interactions. It increases with primary energy $E_0$ as $(N_\mu)_{normal}\sim E_0^\alpha$, with $\alpha<1$. 
The number of \emph{photo-produced muons} $(N_\mu)_{\gamma\to\mu}$ reflects the number of photons in air showers and is nearly proportional to the shower size at maximum  $(N_\mu)_{\gamma\to\mu}\sim (N_e)_{max}\sim E_0$.
Therefore, the fraction of photo-produced muons to normal muons increases with energy: $\frac{(N_\mu)_{\gamma\to\mu}} {(N_\mu)_{normal}}\sim E_0^{1-\alpha}$.
The number of photo-produced muons depends only on the number of photons at the shower maximum both for $\gamma$-ray and hadronic showers.

When a photon photo-produces in the first interaction the shower has a normal hadronic muon content and the cascade is indistinguishable from a proton-induced shower. 

We can estimate the number of muons in a photon-induced shower extending the Heitler model with a simple procedure relatively correct in the shower maximum region \cite{halzen1990,aharonian1991}.
In a shower array the energy threshold of muons typically detected is $\approx$GeV.
According to the Heitler's toy model, in a shower produced by a photon with energy $E_0$, after $t$ radiation lengths, we have a particle cascade which has evolved into $N$ = 2$^t$ particles of equal energy $E$ = $E_0/N$, of which 1/3 are photons. The total number of particles is given by 2 times the number of secondaries at level $t$, the additional factor of 2 taking into account the particles produced in the previous layers.

The number of secondary particles of energy greater than $E_{th}$ has a maximum at a thickness of 1/ln 2 $\cdot$ ln($E_0/E_{th})$ radiation lengths \cite{brossi} (the number of particles at the maximum is $N_{max}\sim (E_0/E_{th}$) ).
A fraction of about 0.6 of the photon energy is transferred to the muons through the production and decay of charged pions. Therefore, the energy of the parent photons able to produce muons relevant to the detection with air shower arrays is $\mathcal{O}(2\>\text{GeV})$.
Therefore, the number of splittings to reach the 2 GeV level is $n\propto$ ln$(E_0/2\>\text{GeV})$.

As an example, in a 100 TeV $\gamma$-induced shower we have that after $t=\frac{1}{\text{ln}\>2}\cdot$ ln$(\frac{100\>\text{TeV}}{2\>\text{GeV}})\approx$15.6 radiation lengths (about 577 g/cm$^2$) $2^t=50,000$ secondaries are produced with an average energy of order of 2 GeV. 
This means that the layer $t\sim 15$ is the last able to produce detectable GeV muons.

The number of muons originating after $t$ layers is $N_\mu= N_\gamma^{TOT}\cdot R_\gamma = 2\cdot N_\gamma(t=15.6)\cdot R_\gamma$ = $2\cdot\frac{1}{3}\cdot 50,000\cdot 3\times 10^{-3}$, where $R_\gamma=\frac{\sigma(\gamma\to\pi)}{\sigma(\gamma\to e^+e^-)}\simeq 3\times 10^{-3}$ is the ratio of the pion PhP cross section to the $e^+e^-$ pair production.
Therefore, the estimated number of muons is $N_\mu\approx 2\cdot 50\approx 100$ in agreement with MonteCarlo simulations. 

Unless R$_\gamma$ exhibits some unexpected energy dependence, the above results are inescapable as the bulk of the muons originates in the last layers where photo-production of pions is exclusively by $\gamma$-rays with energies measured in accelerator experiments.

\begin{figure}[h]
\centering
\includegraphics[width=0.8\textwidth]{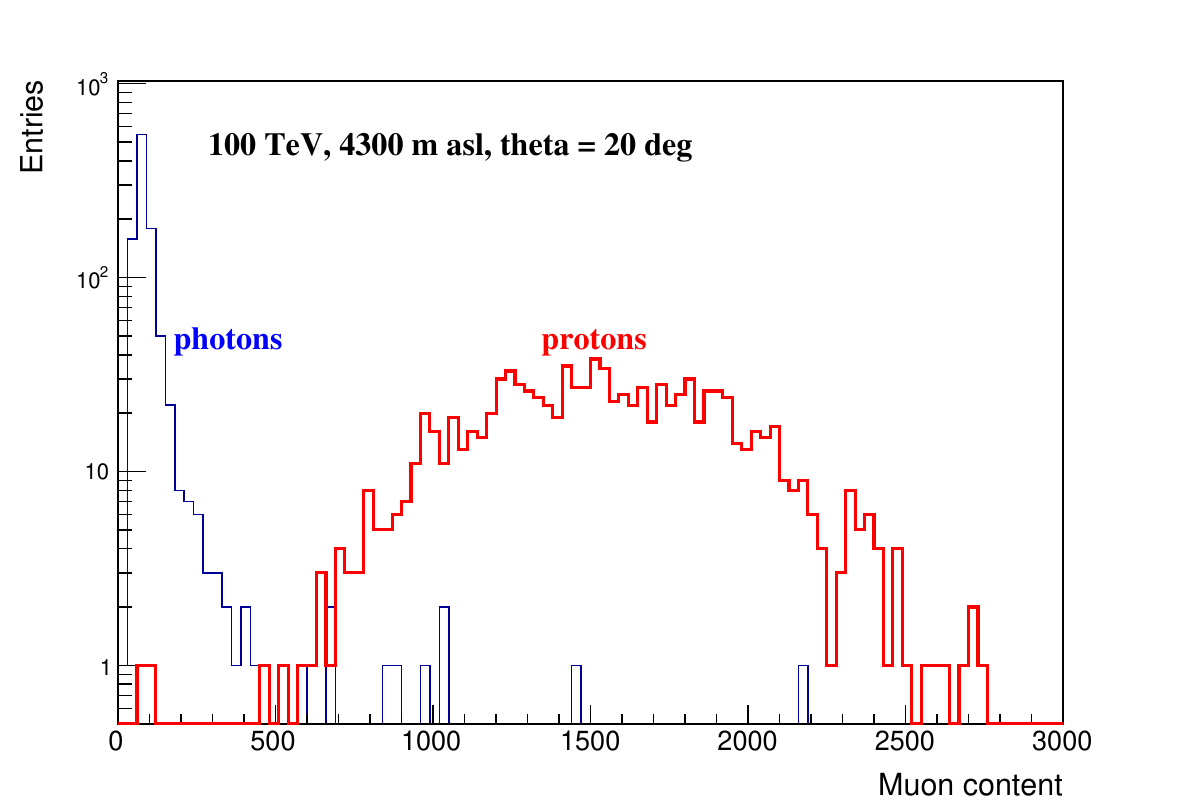}
\caption{Muon distributions in a 100 TeV $\gamma$- and proton-induced showers sampled at 4300 m asl at a zenith angle of 20$^{\circ}$.}
\label{fig:mufluct}       
\end{figure}

In Fig. \ref{fig:mufluct} the muons distributions for 100 TeV gamma- and proton-induced showers sampled at 4300 m asl are shown. The average muon content is $<N_\mu>$= 1543 for proton-induced events and $<N_\mu>$= 97 for $\gamma$-induced showers.

If the first interaction of the primary photon is hadronic the shower is indistinguishable from a normal proton-induced shower. 
As can be seen from the figure, the tail of the muon distribution of $\gamma$-induced showers contains such cascades with the muon content of a hadronic shower. 
These events limit the sensitivity of the ‘muon poor’ technique, but are important to study characteristics of photo-nuclear interactions at high energy.
For a given size N$_e$, the fluctuations in the number of muons in $\gamma$-induced showers are larger than in showers induced by charged cosmic rays because of the competition at each stage of the shower development between the photo-production and pair production cross sections.

We also note some proton-induced events with a very small muon content. At high altitude deep first interactions of the protons are the dominant source of muon-poor hadron showers, at lower altitudes fluctuations towards $\pi^0$-rich showers are the main responsible.
%
%

The detection of $\gamma$-induced events with large muon content is crucial to study the pion photo-production due to high energy photons. 
But these events are typically rejected by background selection cuts based on the muon number and must be carefully evaluated.
Different criteria to select gamma-induced events with large muon size are under study.

\section{Conclusion}

The recent observation of more than 40 gamma sources above 100 TeV by LHAASO \cite{lhaaso} opened a new window in gamma-ray astronomy. The detection of more than 4000 photons per year above 100 TeV and about 20 per year above the PeV in a nearly background-free regime allows for the first time the study of the characteristics of gamma-induced showers and to compare with MonteCarlo simulations. Allows also the study of the hadron nature of photons measuring for the first time the pion photo-production cross section at energies not accessed yet at accelerators. On the other hand, the study of this cross section at energies already investigated at HERA provides a check of the event selection in high energy gamma-ray astronomy.

The construction of the new array SWGO in the Southern Hemisphere \cite{swgo}, where the existence of Super-PeVatrons emitting gammas up to 10 PeV is expected, will provide further sample of pure high energy photons thus extending these studies at higher energies.

\end{document}